\documentclass[aps,prd,twocolumn,nofootinbib]{revtex4-1}
\usepackage{epsfig}
\usepackage{amsfonts}
\usepackage[colorlinks,linkcolor=blue,anchorcolor=blue,citecolor=blue,urlcolor=blue,breaklinks=true]{hyperref}
\usepackage{graphics}
\usepackage{multirow}
\usepackage{color}
\usepackage{amsmath}
\usepackage{ulem}
\newcommand{\MeV}{\mathrm{MeV}}
\newcommand{\Mev}{\mathrm{MeV}}
\newcommand{\btau}{\boldsymbol{\tau}}
\begin{document}
\author{Ya Lu$^{1}$}
\author{Yi-Lun Du$^{1}$}
\author{Zhu-Fang Cui$^{1,3}$}
\author{Hong-Shi Zong$^{1,2,3,}$}\email{Email:zonghs@nju.edu.cn}

\address{$^{1}$Department of Physics, Nanjing University, Nanjing 210093, China}
\address{$^{2}$Joint Center for Particle, Nuclear Physics and Cosmology, Nanjing 210093, China}
\address{$^{3}$ State Key Laboratory of Theoretical Physics, Institute of Theoretical Physics, CAS, Beijing, 100190, China}

\title{Critical behaviors near the (tri-)critical end point of QCD within the NJL model}
\begin{abstract}
We investigate the dynamical chiral symmetry breaking and its restoration at finite density and temperature within the two-flavor Nambu-Jona-Lasinio model, and mainly focus on the critical behaviors near the critical end point (CEP) and tricritical point (TCP) of quantum chromodynamics. The multi-solution region of the Nambu and Wigner ones is determined in the phase diagram for the massive and massless current quark, respectively. We use the various susceptibilities to locate the CEP/TCP and then extract the critical exponents near them. Our calculations reveal that the various susceptibilities share the same critical behaviors for the physical current quark mass, while they show different features in the chiral limit.

\bigskip

\end{abstract}
\maketitle

\section{INTRODUCTION}

Quantum Chromodynamics (QCD) is often viewed as the basic theory of strong interactions, whose degrees of freedom include quarks and gluons. As a non-Abelian gauge field theory, QCD exhibits two important features: asymptotic freedom and color confinement. It works well in the large-momentum-transfer processes through the perturbative techniques due to the feature of asymptotic freedom. However, in the small-momentum-transfer processes, the coupling constant becomes so strong that problems have to be treated by many non-perturbative  methods, such as lattice QCD~\cite{Braguta:2015zta},  Nambu-Jona-Lasinio (NJL) model~\cite{RevModPhys.64.649,Buballa2005205}, Dyson-Schwinger equations~\cite{PPNP.77.1-69} and so on.

It is believed that the strongly interacting matter undergoes some phase transition and the chiral symmetry is partially restored at high temperature ($T$) and/or quark chemical potential ($\mu$), from the hadronic matter to the quark-gluon plasma (QGP). This phase transition is expected to be produced in on-going heavy-ion collision experiments, such as the BNL Relativistic Heavy-Ion Collider (RHIC) and the CERN Large Hadron Collider (LHC)~\cite{PhysRevLett.112.032302,*PhysRevLett.113.092301,PhysRevSTAB.17.081004,alice2014performance}. In addition, high density QCD matter at low temperature is anticipated in the compact stars physics~\cite{Klahn2007170,PhysRevD.72.065020,PhysRevD.86.114028}.

Lattice QCD works well for low $\mu$ and finite $T$. However, the lattice simulations have trouble in studying the QCD phase transitions with finite $\mu$ due to the severe fermion sign problem. Therefore simpler and mathematically tractable models which respect the essential symmetries of the QCD are necessary. NJL model captures fundamental features of QCD itself, such as the dynamical chiral symmetry breaking in the vacuum, and can therefore yield profound insight into the critical behaviors associated with chiral symmetry.

Many effective theories~\cite{alford1998qcd,PhysRevD.58.096007,PhysRevD.77.096001} of QCD predict the existence of the tricritical point (TCP) and the critical end point (CEP) in the QCD $T-\mu$ phase diagram in massless and massive quark cases, respectively. TCP is a point where three coexisting phases become identical simultaneously. The chiral phase transition is of first-order at high $\mu$ and low $T$, then turns to be second-order after the TCP.  The second-order transition line is expected to be in the universality class of three dimensional $O(4)$ symmetric spin model, also called $O(4)$ line. With the current quark masses increasing from zero to the physical masses, TCP emerges to CEP at which the first-order phase transition line ends from higher $\mu$ toward lower $\mu$. Meanwhile, the second-order phase transition turns into a smooth crossover at lower $\mu$ and higher $T$ in the phase diagram. Different from the $O(4)$ line, the CEP is expected to be in the $Z_2$ universality class~\cite{schmidt2013baryon}, this is just one of several possible scenarios for QCD.


Both at the TCP and CEP, second-order phase transitions are expected to occur. Therefore, the values of critical exponents will be an interesting and important question. The critical exponents in mean field approximation have been predicted, by expanding the Ginzburg-Landau thermodynamic potential to the order parameter~\cite{PhysRevD.67.014028}. Therefore it is important to verify these universality arguments from the study of specific effective theory of QCD, for this purpose we investigate the critical behaviors of the QCD by using the NJL model for two quark flavors.

The rest of this paper is organized as follows: In Sec. \ref{njl} a mean field description of the Nambu-Jona-Lasinio model with two quark flavors as an effective realization of the low-energy sector of QCD is presented. In Sec. \ref{tcpt} we display the multi-solution region of the Nambu and Wigner ones and analyze the behaviors of the effective quark mass and the quark number density in the $T-\mu$ plane for massive and massless current quark, respectively. In Sec. \ref{sace} we analyze the behaviors of some susceptibilities in the $T-\mu$ plane and determine the location of the CEP (or TCP). Then critical exponents on the CEP are calculated and compared with those of the TCP. Finally, in Sec. \ref{sac} we will summarize our results and give the conclusions.

\section{THE NAMBU-JONA-LASINIO MODEL}\label{njl}

The Lagrangian of the NJL model~\cite{RevModPhys.64.649,Buballa2005205,Masayuki1989668} is written as:
\begin{eqnarray}
\mathcal{L}_{NJL}&=&\bar{\psi}(i{\not\!\partial}-m)\psi+\mathcal{L}_I\nonumber\\
&=&\bar{\psi}(i{\not\!\partial}-m)\psi+g[(\bar{\psi}\psi)^2+(\bar{\psi}i\gamma_5{\btau}\psi)^2].
\end{eqnarray}
The column vector $\psi=(u,d)$ represents the quark fields, where we take the number of flavors $N_f=2$, and the number of colors $N_c=3$ throughout this work. $g$ is the effective coupling constant of the four fermion interaction, and $\btau$ are the Pauli matrices in the SU$(2)$ flavor space. The Fierz transformation of $\mathcal{L}_I$ is
\begin{eqnarray}
\mathcal{L}_{IF}=&&\frac{g}{8N_c}[2(\bar{\psi}\psi)^2+2(\bar{\psi}i\gamma_5\btau\psi)^2-2(\bar{\psi}\btau\psi)^2\nonumber\\
                 &&-2(\bar{\psi}i\gamma_5\btau\psi)^2-4(\bar{\psi}\gamma^{\mu}\psi)^2\nonumber\\
                 &&-4(\bar{\psi}i\gamma^{\mu}\gamma_5\psi)^2+(\bar{\psi}\sigma_{\mu\nu}\psi)^2-(\bar{\psi}
    \sigma_{\mu\nu}\btau\psi)^2].
\end{eqnarray}
We adopt only  $\bar{\psi}\psi$ and $\bar{\psi}\gamma_0\psi$ as mean fields on account of symmetry properties of the vacuum  at finite density, and ignore other terms. Then the interaction term is written
\begin{eqnarray}
\mathcal{L}_{IMF}=&&2G\langle\bar{\psi}\psi\rangle\bar{\psi}\psi-\frac{g}{N_c}\langle\bar{\psi}\gamma_0\psi\rangle\bar{\psi}\gamma_0\psi\nonumber\\
&&-G{\langle\bar{\psi}\psi\rangle}^2+\frac{g}{2N_c}{\langle\bar{\psi}\gamma_0\psi\rangle}^2,
\end{eqnarray}
where $G=(4N_c+1)g/{4N_c}$ is the renormalized coupling constant, and we define the thermal expectation value of an operator $\mathbb{O}$ by
\begin{equation}
\langle\mathbb{O}\rangle=\frac{\mathrm{Tr}\ \mathbb{O}\ \mathrm{e}^{-\beta(\mathcal{H}-\mu \mathcal{N})}}{\mathrm{Tr}\ \mathrm{e}^{-\beta(\mathcal{H}-\mu \mathcal{N})}}.
\end{equation}

By using Fierz transformation and mean field approximation we get the Hamiltonian density:
\begin{equation}\label{eqn:H}
\mathcal{H}=-i \bar{\psi}  \gamma \cdot \bigtriangledown \psi+ M \bar{\psi} \psi+\frac{g}{N_c}\sigma_2\mathcal{N}+G\sigma_1^2-\frac{g}{2N_c}\sigma_2^2,
\end{equation}
where $\mathcal{N}=\bar{\psi}\gamma_0\psi=\psi^\dag\psi$ is the quark number density.
$M$ is the constituent mass (the effective quark mass):
\begin{equation}\label{eqn:M}
M=m-2G{\sigma_1}(M,\mu_r).
\end{equation}
$\sigma_1$ and $\sigma_2$ are defined as follows: $\sigma_1=\langle\bar{\psi}\psi\rangle$, $\sigma_2=\langle\bar{\psi}\gamma_0\psi\rangle$. We also introduce the renormalized chemical potential:
\begin{equation}\label{eqn:mur}
{\mu}_r=\mu-\frac{g}{N_c}\sigma_2(M,\mu_r).
\end{equation}

To calculate $M$, $\mu_r$ self-consistently by Eqs. (\ref{eqn:M}) and (\ref{eqn:mur}) we use the formalism of the thermal Green function in the real time~\cite{PhysRevD.9.3320}. So the Green function of a free fermion at temperature $T$ and chemical potential $\mu$ is given as
\begin{eqnarray}
G(p;T,\mu)=&&(\not\!p+M)\{\frac{1}{p^2-M^2+i\epsilon}+2\pi i\delta(p^2-M^2)\nonumber\\
            &&\times[\theta(p^0)n(p,\mu)+\theta(-p^0)m(p,\mu)]\},
\end{eqnarray}
with

\begin{equation}
n(p,\mu)=\frac{1}{1+\mathrm{e}^{\beta(E-\mu)}},\ \ \ \ \ m(p,\mu)=\frac{1}{1+\mathrm{e}^{\beta(E+\mu)}},
\end{equation}
$\beta=1/T$, and $E=\sqrt{M^2+p^2}$. Using the Green function we get $\sigma_1$ and $\sigma_2$ as follows:
\begin{eqnarray}
\sigma_1&&=-M\frac{N_cN_f}{\pi^2}\!\!\int_0^\Lambda\!\frac{p^2}{E}[1-n(p,\mu_r)-m(p,\mu_r)]dp, \label{eqn:sigma1}\\
\sigma_2&&=\frac{N_cN_f}{\pi^2}\int_0^\Lambda p^2[n(p,\mu_r)-m(p,\mu_r)]dp.              \label{eqn:sigma2}
\end{eqnarray}

In this paper, we employ the parameter set used by Hastuda and  Kunihiro~\cite{Hatsuda1994221}: $m=5.5~{\MeV}$, $g=5.074\times10^{-6}~{\MeV}^{-2}$, $\Lambda=631 ~{\MeV}$, which are determined by fitting the pion mass $m_\pi=138~{\MeV}$, the pion decay constant $f_\pi=93.1~{\MeV}$. For the quark condensates we obtain ${\langle\bar{\psi}\psi\rangle}^{1/3}=-331 ~{\MeV}$.

\begin{figure}
\includegraphics[width=0.4\textwidth]{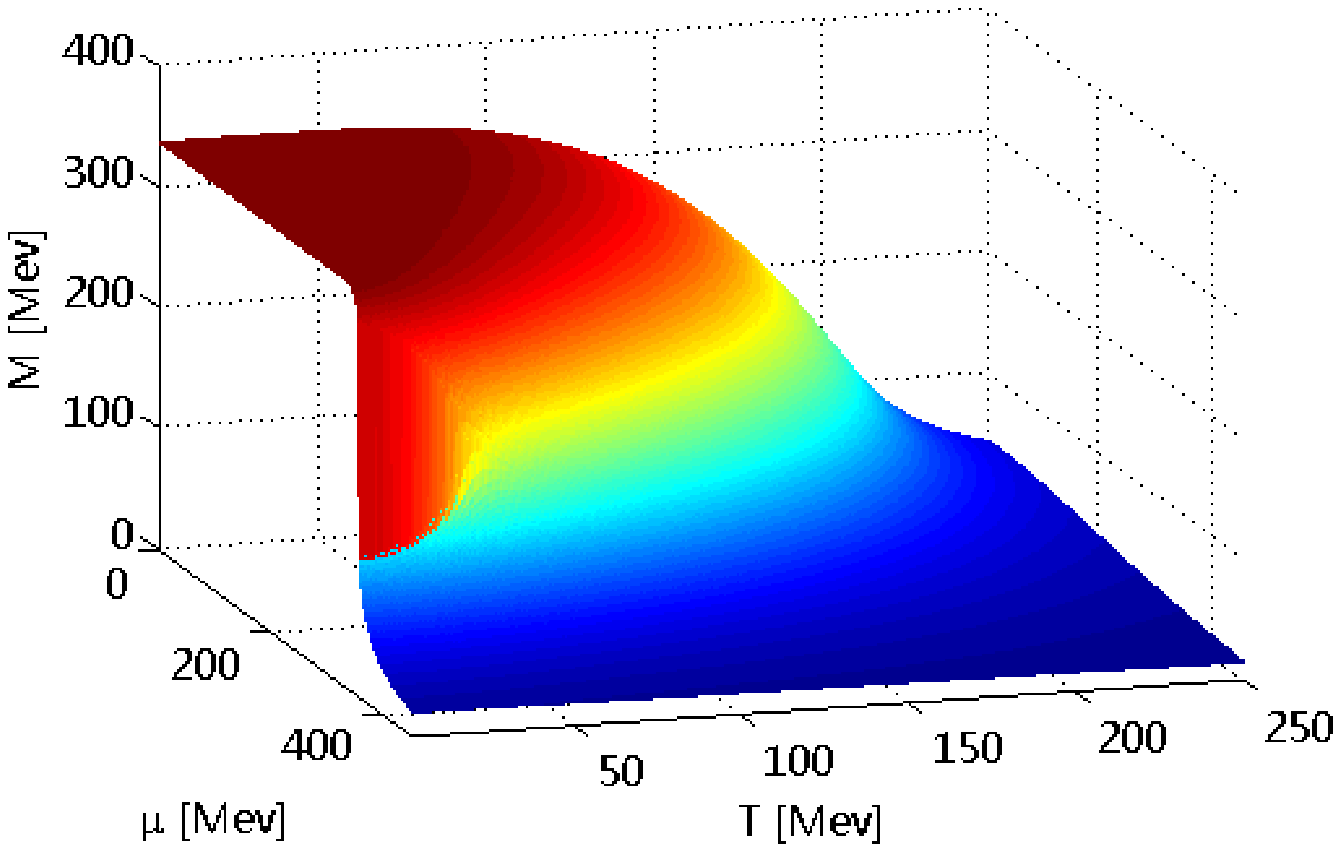}
\includegraphics[width=0.4\textwidth]{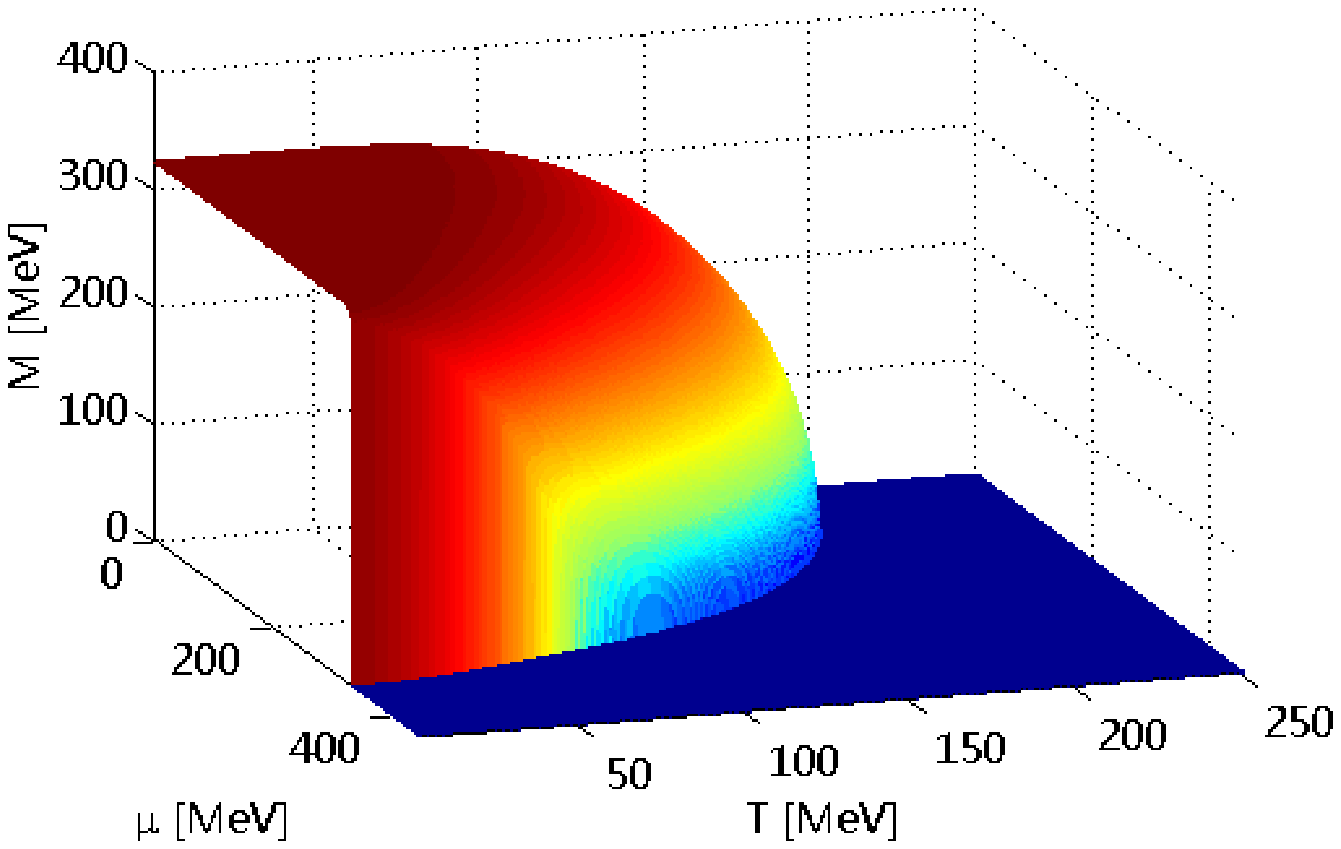}
\caption{(color online). The effective quark mass $M$ in the $T-\mu$ plane. The upper panel corresponds to the case of finite quark mass and the lower panel to the chiral limit.}\label{m}
\end{figure}

Now one can get the effective quark mass and the quark number density for each temperature and chemical potential by solving self-consistent Eqs. (\ref{eqn:M}), (\ref{eqn:mur}), (\ref{eqn:sigma1}), and (\ref{eqn:sigma2}). In the multi-solution region of Nambu solution (chiral symmetry broken) and Wigner solution (chiral symmetry partially restored), the solution which minimizes the thermodynamical potential density $\omega$ is stable. From Eq.~(\ref{eqn:H}) the thermodynamical potential density $\omega$ is given by
\begin{eqnarray}
\omega=&&\frac{-T}{V}\log [ \mathrm{Tr}\ \mathrm{e}^{\beta(\mu\mathcal{N}-\mathcal{H})}]\nonumber\\
      =&&G\sigma_1^2-\frac{g}{2N_c}\sigma_2^2-\frac{N_c N_f}{\pi^2}\int_0^\Lambda p^2 E dp\nonumber\\
       &&-\frac{N_c N_f}{\pi^2} T \int_0^\Lambda p^2 \{\log[1+\mathrm{e}^{\beta(\mu_r-E)}]\nonumber\\
       &&+\log[1+\mathrm{e}^{\beta(-\mu_r-E)}]\} dp.
\end{eqnarray}
Having taken this into consideration, we obtain the effective quark mass in the $T-\mu$ plane in Fig.~\ref{m} in the massive and massless current quark case, respectively.

\bigskip
\section{THE CHIRAL PHASE TRANSITION}\label{tcpt}

\begin{figure}
\includegraphics[width=0.47\textwidth]{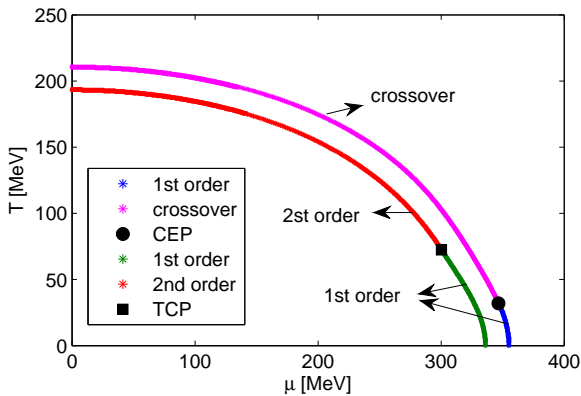}
\caption{(color online). Phase diagram in the two-flavor NJL model. The green (blue in the chiral limit) points represent the first-order phase transition, the pink points the crossover transition, and the red points the second-order phase transition.}\label{phase}
\end{figure}

The phase diagram of QCD within the two-flavor NJL model is shown in Fig.~\ref{phase}. It should be noticed that there are multi-solution around the first-order transition line~\cite{CPL.22.3036,PhysRevC.75.015201,Williams2007167,PhysRevD.86.114001,cui2013wigner,*cui2014wigner,jiang2013chiral,*shi2014locate}. The multi-solution of two solutions indicates the competition of the two corresponding phases, which is a characteristic of the first-order phase transition. Such a consideration is very important for determining the first-order transition line and locating the CEP. The boundaries of multi-solution region have been determined in the Refs.~\cite{deForcrand200662,PhysRevLett.106.172301} by the approach of NJL model and the Dyson-Schwinger equations, respectively.

\begin{figure}
\includegraphics[width=0.47\textwidth]{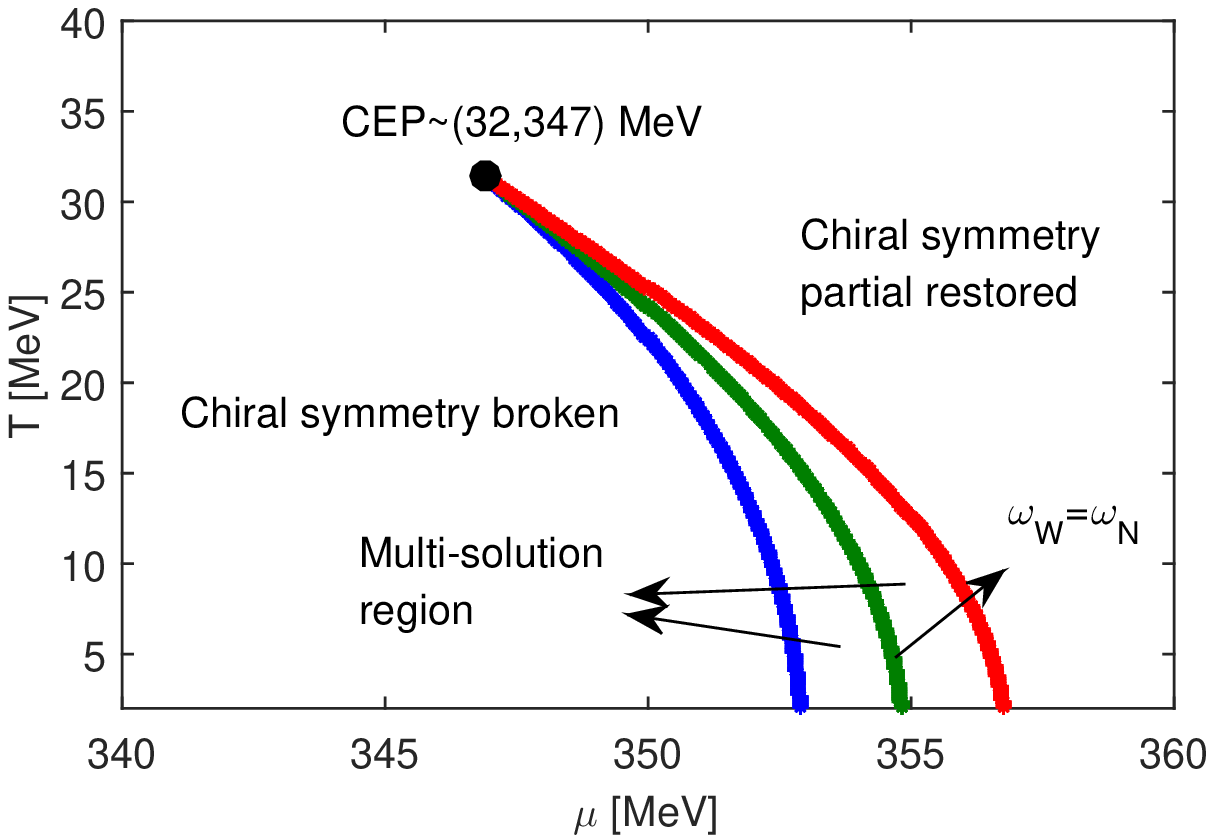}
\includegraphics[width=0.47\textwidth]{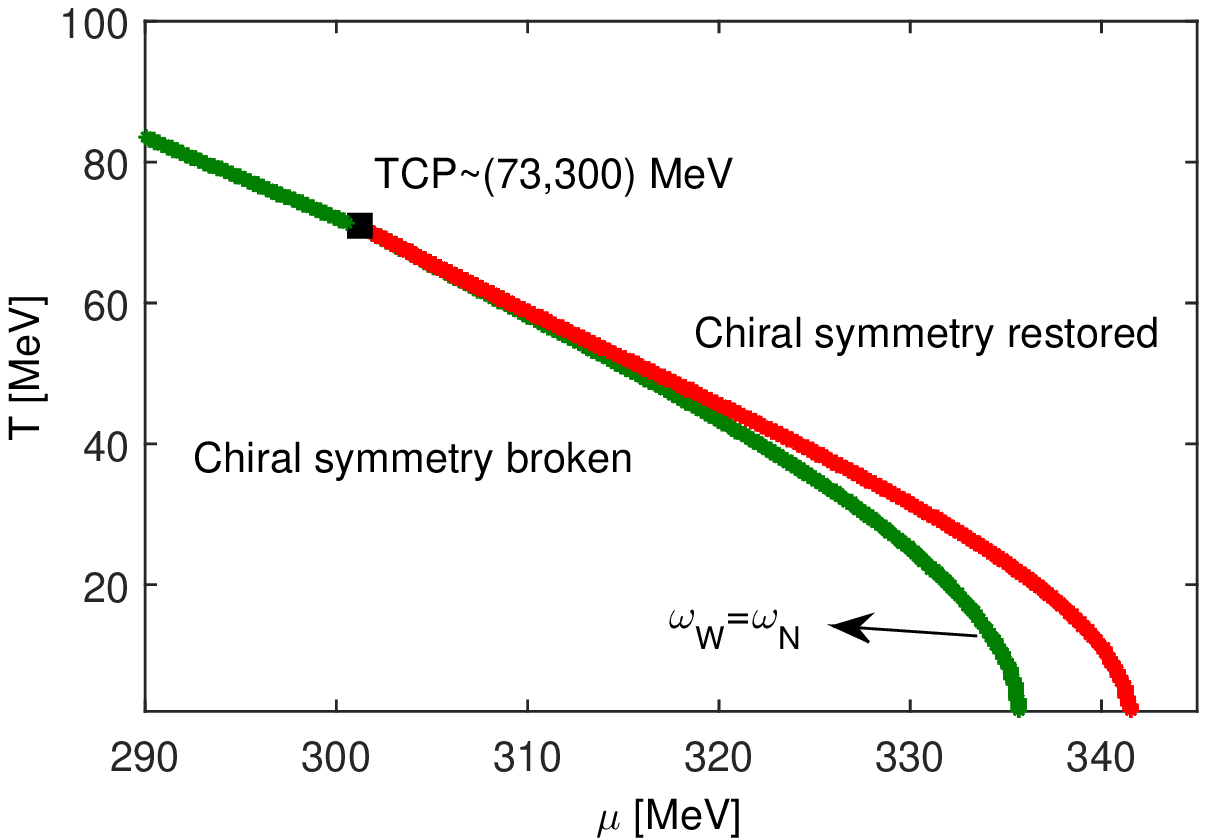}
\caption{(color online). Phase diagram in the temperature-chemical potential plane for strongly interacting quarks around the CEP/TCP. The upper panel corresponds to the case of finite quark mass and the lower panel to the chiral limit.}\label{nw}
\end{figure}

We display the multi-solution region in Fig.~\ref{nw}, for massive and massless current quark, respectively.  For the massive current quark case, the borders of the multi-solution region are marked by the blue and red lines in Fig.~\ref{nw}. The blue line is where the Wigner solution begins to appear, while the red one is where the Nambu solution disappears. The multi-solution region is split into two parts by the first-order transition line (the green line in Fig.~\ref{nw}). On this line, the thermodynamical potential densities of the Nambu phase ($\omega_N$) and Wigner phase ($\omega_W$) are degenerate. The corresponding Nambu phase is stable and the Wigner phase is metastable in the left part, while the Wigner phase is stabler in the right than the Nambu one.  With $T$ increasing, the multi-solution region tapers off on approaching the CEP ($T^\mathrm{CEP} \sim 32~{\Mev}$, $\mu^\mathrm{CEP} \sim 347~{\Mev}$).

For the massless current quark case, the massless quark is the trivial solution for each temperature and chemical potential. The first-order phase transition ends at the TCP ($T^\mathrm{TCP} \sim 73~{\Mev}$, $\mu^\mathrm{TCP} \sim 300~{\Mev}$), In the domain where $T>T^\mathrm{TCP}$ and $\mu<\mu^\mathrm{TCP}$, a second-order phase transition instead of the crossover transition occurs.

\bigskip
\section{SUSCEPTIBILITIES AND CRITICAL EXPONENTS}\label{sace}

\subsection{Susceptibilities}\label{sus1}

As the linear response of the physical system to some external field,  susceptibility is often measured to study the properties of the related system. Therefore the studies of various susceptibilities are very important on the theoretical side, which are widely used to study the phase transitions of strongly interacting matter~\cite{cui2015progress}. So here we introduce several kinds of susceptibilities: the chiral susceptibility $\chi_s$, the quark number susceptibility $\chi_q$, the thermal susceptibility $\chi_T$, the vector-scalar susceptibility $\chi_{vs}$, the susceptibilities $\chi_m$ and $\chi_n$ (for mathematical convenience) defined as follows~\cite{PhysRevD.88.114019}:

\begin{eqnarray}
&&\chi_s=-\frac{\partial \langle\bar{\psi}\psi\rangle}{\partial m},\hspace{8mm}
\chi_q=\frac{\partial \langle{\psi}^\dag \psi\rangle}{\partial\mu},\nonumber\\
&&\chi_T=\frac{\partial \langle\bar{\psi}\psi\rangle}{\partial T},\hspace{10mm}
\chi_{vs}=\frac{\partial \langle\bar{\psi}\psi\rangle}{\partial\mu},\nonumber\\
&&\chi_m=-\frac{\partial \langle{\psi}^\dag \psi\rangle}{\partial m},\hspace{6mm}
\chi_n=\frac{\partial \langle{\psi}^\dag \psi\rangle}{\partial T}.
\end{eqnarray}
From the viewpoint of statistical mechanics it is easy to find that $\chi_m$ equals $\chi_{vs}$: $\chi_m=\chi_{vs}=\frac{T}{V}\frac{\partial^2}{\partial m\partial\mu}\ln{Z}$, where $Z$ is the QCD partition function.

\begin{figure}
\includegraphics[width=0.4\textwidth]{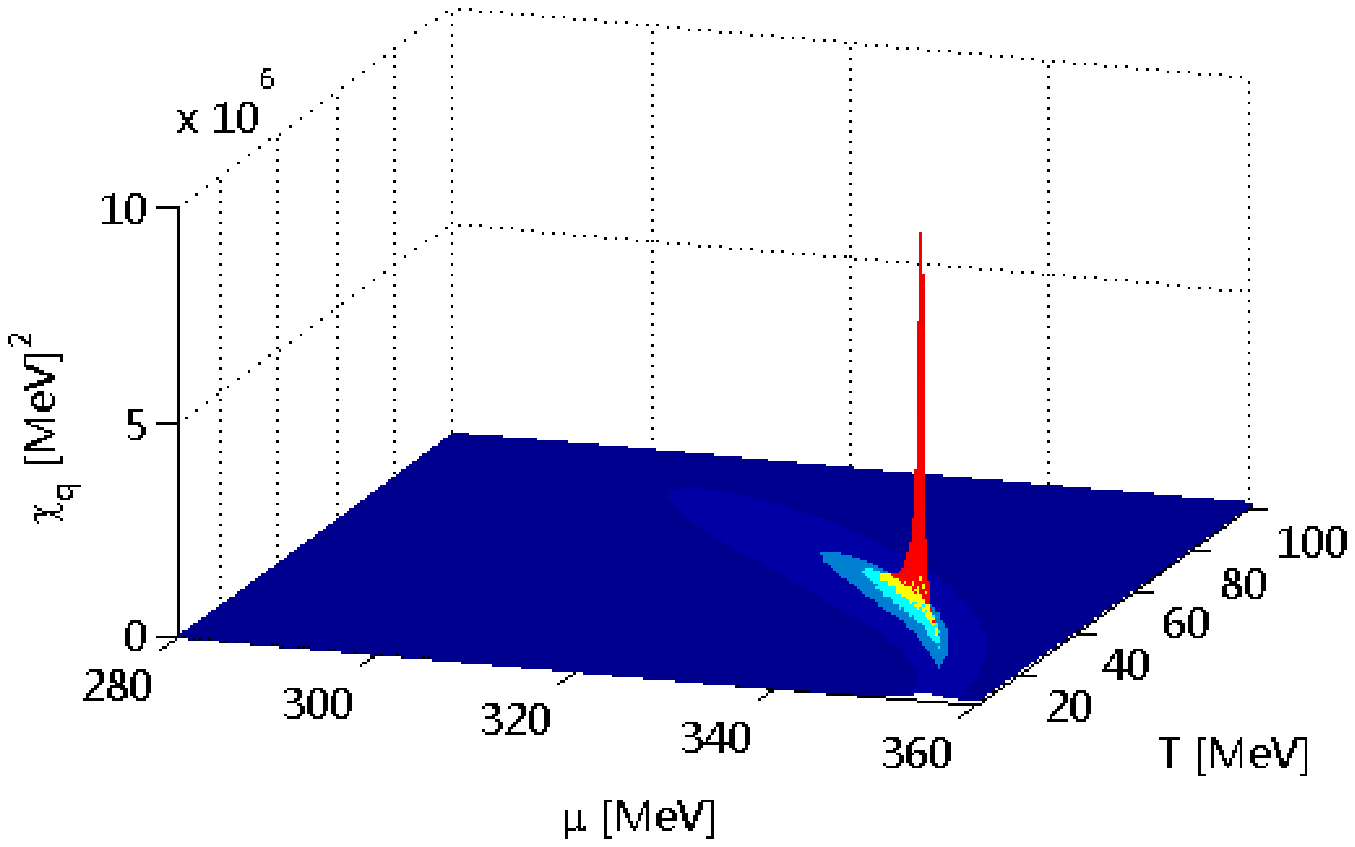}
\includegraphics[width=0.4\textwidth]{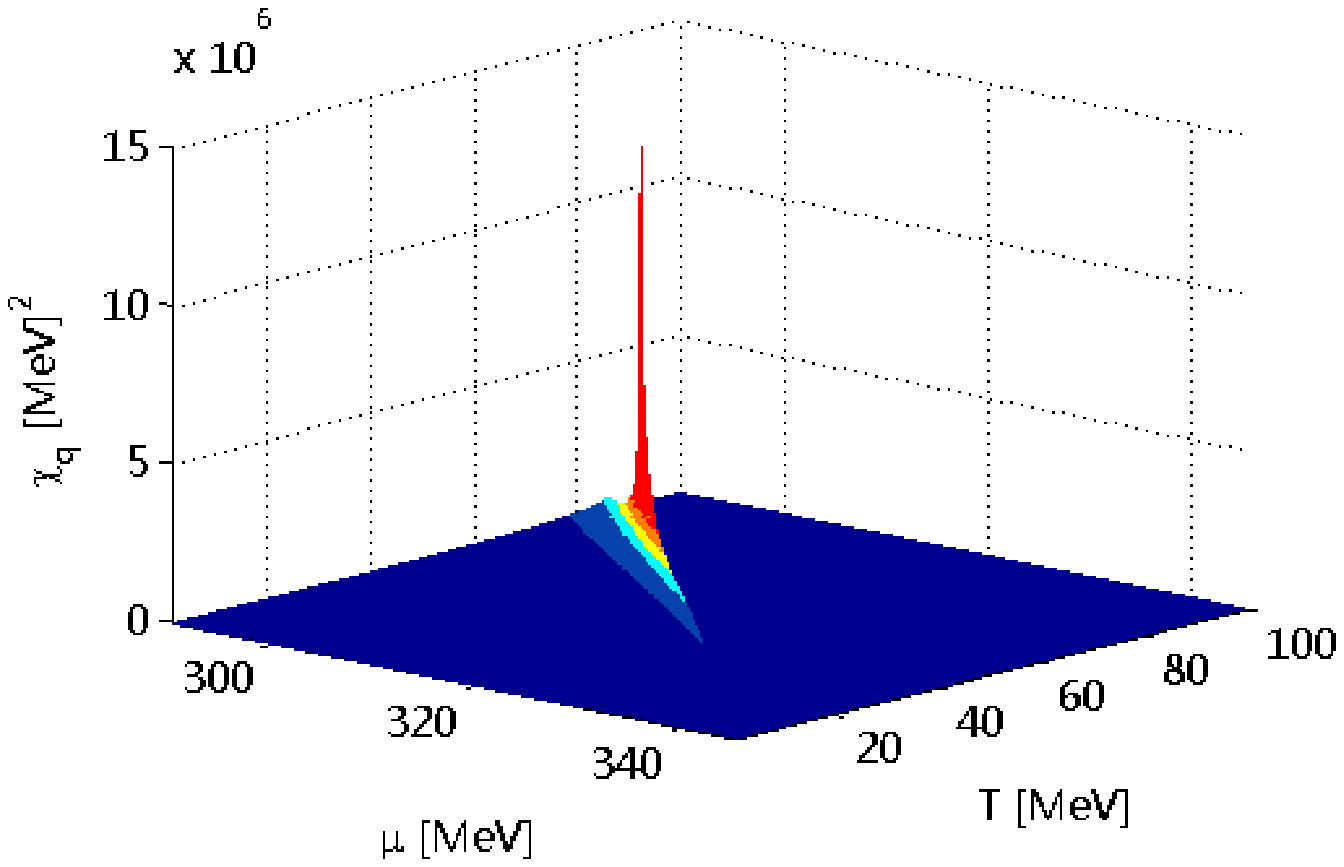}
\caption{(color online). 3D plot for the quark number susceptibility $\chi_{q}$ around the first-order transition line and CEP/TCP in the $T-\mu$ plane. The upper panel corresponds to the case of finite quark mass and the lower panel to the chiral limit.}\label{chiq}
\end{figure}

Here we only display the quark number susceptibility in the $T-\mu$ plane in Fig.~\ref{chiq} as an example. In the case of finite current quark mass, all susceptibilities, i.e., $\chi_s$, $\chi_q$, $\chi_T$, $\chi_{vs}$, $\chi_m$ and $\chi_n$ change discontinuously on the first-order phase transition line, and all diverge at the CEP. Then the susceptibilities are continuous in the crossover region.

In the chiral limit, there is a second-order phase transition line instead of the crossover, and the susceptibilities always change discontinuously no matter on the first- or second-order phase transition line. All those susceptibilities diverge at the same point, where the first-order transition line ends. Therefore they all suggest the same TCP. The susceptibilities show different features on the second-order phase transition line: $\chi_s$, $\chi_T$ and $\chi_m$ diverge, while $\chi_q$ and $\chi_n$ remain finite. The different features of the various susceptibilities result in the differences of the corresponding critical exponents calculated in Sec. \ref{ce1}.

\subsection{Critical exponents}\label{ce1}

The susceptibility in the vicinity of the CEP/TCP diverges with an index, the so-called critical exponent.  The values of these exponents are completely dependent on the dimension of space and components of the order parameter instead of the details of the microscopic dynamics. Hence all theories can be categorized into a much smaller number of universality classes~\cite{PhysRevC.58.1758}. However, the critical exponents we have obtained are mean field values, due to the mean field approximation employed in this work.

The strength of the divergence is dependent on the path, along which we approach the critical points~\cite{PhysRevLett.24.715}. We mark the critical exponents by $\gamma^\mathrm{CEP}$, $\gamma^\mathrm{TCP}$  for the massive and massless current quark, respectively. As is well known, the critical exponents of quark number susceptibility are expected $\gamma^\mathrm{CEP}_q=\gamma^\mathrm{TCP}_q=1$ for the path asymptotically parallel to the first-order transition line. For other paths the critical exponents are expected to be $\gamma^\mathrm{CEP}_q=2/3$ and $\gamma^\mathrm{TCP}_q=1/2$~\cite{PhysRevD.67.014028}.

\begin{figure}
\includegraphics[width=0.47\textwidth]{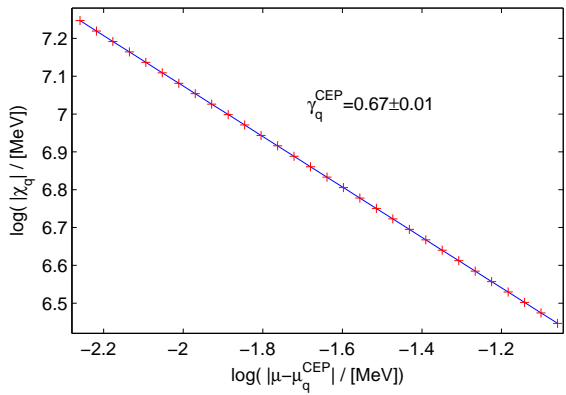}
\caption{(color online). The logarithm value of the quark number susceptibility as a function of $\log|\mu-\mu^\mathrm{CEP}|$ at the fixed temperature $T^\mathrm{CEP}$.}\label{fit}
\end{figure}

Here we calculate four directions of the critical exponents of susceptibilities: the path from lower $\mu$ toward $\mu^\mathrm{C}$ (represents $\mu^\mathrm{CEP}$ or $\mu^\mathrm{TCP}$) with the temperature fixed at $T^\mathrm{C}$ (represents $T^\mathrm{CEP}$ or $T^\mathrm{TCP}$), we mark this path by $\rightarrow$. Analogously, other arrows $\leftarrow$, $\uparrow$, $\downarrow$ represent the path from higher $\mu$ toward $\mu^\mathrm{C}$, the path from lower and higher $T$ toward $T^\mathrm{C}$, respectively. Using the linear logarithmic fit we obtain

\begin{eqnarray}
 &&\ln{\chi}=-\gamma \ln|T-T^\mathrm{C}|+c_1,\\
 &&\ln{\chi}=-\gamma \ln|\mu\hspace{0.5mm}-\hspace{0.5mm}\mu^\mathrm{C}|+c_2,
\end{eqnarray}
for the directions parallel to the $\mu$ axis and $T$ axis, respectively. $c_1$, $c_2$ are constants and $\gamma$ is the critical exponent. The fitting procedure of the critical exponent for quark number susceptibility for the direction $\rightarrow$, is shown in Fig.~\ref{fit}.

For the case of finite current quark mass, the order parameter only carries one component---sigma ($\sigma$), which is the only field becomes massless at the CEP. As is widely accepted the phase transition falls into the three-dimensional (3D) Ising model as the liquid-gas phase transition~\cite{berges1999color,Phys.Rev.D.58.096007,PhysRevLett.81.4816}. Our results presented in TABLE.~\ref{tabcep} show that, the critical exponents of $\gamma_{s}^\mathrm{CEP}$ and $\gamma_{q}^\mathrm{CEP}$ agree with the mean field prediction 2/3 for all four different directions to approach the CEP. As far as we know, there is no discussion on the critical exponents of $\gamma_{T}^\mathrm{CEP}$, $\gamma_m^\mathrm{CEP}$ and $\gamma_n^\mathrm{CEP}$ in previous literatures, our results imply that they are also 2/3. Because of the finite current quark mass and quark chemical potential, the critical fluctuation mixes in the quark number density fluctuation and chiral condensate fluctuation as well as in the thermal fluctuation, and therefore all the relevant susceptibilities show the same critical behavior.  Furthermore, the ``t'' and ``H'' fields of the 3D Ising model are mapped as certain linear combinations of $T$ and $\mu$ on the $T -\mu$ plane near the CEP.  The singular behavior of the susceptibilities is insensitive to the way of approaching the CEP. It only depends on the fact whether the approach is tangential or non-tangential to the ``t'' direction. Therefore it is possible that this character holds even beyond the mean field approximation (nevertheless, the real scenario need to be checked in further study).

\begin{table}
\begin{center}
\caption{\label{tabcep}Critical exponents at the CEP.}
\begin{tabular}{c c c c r @{.} l}\hline\hline
Quantity &  \hspace{3mm}Path    &  \hspace{3mm}  Numerical result & \hspace{3mm}MF exponent \\\hline
 \multirow{4}{*}{$\gamma_s^\mathrm{CEP}$}
          &$\rightarrow$   &  0.69$\pm$0.01                       &  \multirow{4}{*}{2/3} \\
         &      $\leftarrow$    &    0.65$\pm$0.01                            \\
         &      $\uparrow$      &    0.69$\pm$0.01                       \\
         &      $\downarrow$    &    0.64$\pm$0.01                       \\\hline

 \multirow{4}{*}{$\gamma_q^\mathrm{CEP}$}
         &$\rightarrow$  &   0.67$\pm$0.01                   & \multirow{4}{*}{2/3}  \\
         &      $\leftarrow$    &    0.67$\pm$0.01                           \\
         &      $\uparrow$      &    0.67$\pm$0.01                             \\
         &      $\downarrow$    &    0.66$\pm$0.01                             \\\hline
 \multirow{4}{*}{$\gamma_T^\mathrm{CEP}$}
         &$\rightarrow$   &  0.64$\pm$0.01                         & \multirow{4}{*}{---}  \\
         &      $\leftarrow$    &    0.69$\pm$0.01                               \\
         &      $\uparrow$      &    0.66$\pm$0.01                               \\
         &      $\downarrow$    &    0.68$\pm$0.01                               \\\hline
 \multirow{4}{*}{$\gamma_m^\mathrm{CEP}$}
         & $\rightarrow$   &  0.68$\pm$0.01                         & \multirow{4}{*}{---} \\
         &      $\leftarrow$    &    0.66$\pm$0.01                              \\
         &      $\uparrow$      &    0.69$\pm$0.01                            \\
         &      $\downarrow$    &    0.65$\pm$0.01                             \\\hline
  \multirow{4}{*}{$\gamma_n^\mathrm{CEP}$}
          &$\rightarrow$   &  0.66$\pm$0.01                         &\multirow{4}{*}{---}  \\
         &      $\leftarrow$    &    0.67$\pm$0.01                               \\
         &      $\uparrow$      &    0.65$\pm$0.01                              \\
         &      $\downarrow$    &    0.68$\pm$0.01                              \\\hline\hline
 \end{tabular}
\end{center}
\end{table}

For the reason that the critical region around the TCP is chopped off in the chirally symmetric phase, we only calculated two directions of the critical exponents, i.e., $\rightarrow$ and $\uparrow$, as listed in TABLE.~\ref{tabtcp} in the chiral limit. We obtain $\gamma^\mathrm{TCP}_s$ and $\gamma^\mathrm{TCP}_q$ which agree with the prediction  $\gamma^\mathrm{TCP}_s=1$~\cite{PhysRevD.77.034024}, and $\gamma^\mathrm{TCP}_q=1/2$~\cite{PhysRevD.77.034024,costa2007qcd,*costa2009qcd}
for both directions to approach the TCP.  We also calculate the critical exponent of the order parameter $\langle\bar{\psi}\psi\rangle$ marked by $\beta$, which agrees with the Ginzburg-Landau effective theory prediction 1/4~\cite{PhysRevD.77.034024}.
In the vicinity of the TCP $\langle\bar{\psi}\psi\rangle \sim |T^\mathrm{TCP}-T|^{1/4}$, so it is reasonable to expect the critical exponents of susceptibility $\chi_T$ is equal to 3/4, since $\chi_T={\partial{\langle\bar{\psi}\psi\rangle}}/{\partial{T}}$. Analogously, the critical exponents of $\chi_m$ is also expected to be 3/4, as $\chi_m=\chi_{vs}={\partial{\langle\bar{\psi}\psi\rangle}}/{\partial{\mu}}$. Our numerical calculations suggest that the critical exponents of $\gamma_T^\mathrm{TCP}$ and $\gamma_m^\mathrm{TCP}$ are 3/4, while $\gamma_n^\mathrm{TCP}=1/2$ .
More studies related to the critical exponents near CEP and TCP can be found in Refs.~\cite{PhysRevC.59.1751,PhysRevD.77.096001,Phys.Rev.D.50.6954,PhysRevD.82.054026,PhysRevD.75.085015,*PhysRevD.85.034027}. Whereas the critical exponents of TCP are classical in three dimensions (since the upper critical dimension is 3), this is not the case for the critical exponents of the CEP. The critical exponents in the universality class of $Z_2$ are substantially different from the mean field exponents. Mean field exponents are classical for D=4, and do not correspond to any D=3 universality class. So it is worth noting that the critical exponents obtained in this paper are the mean field ones, due to the mean field approximation we used in this work.

\begin{figure}
\includegraphics[width=0.47\textwidth]{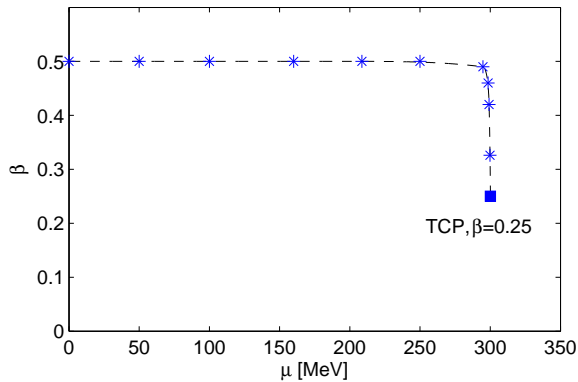}
\caption{(color online). Critical exponents as a function of $\mu$ on the  second-order line.} \label{beta}
\end{figure}

Moveover, in the chiral limit,  $O(4)\sim SU(2)_V \times SU(2)_A$ symmetry is restored on the second-order phase transition line~\cite{Stephanov2005}. That is, the order parameter carry four components---sigma and three pions ($\sigma,\boldsymbol{\pi}$), which all become massless at those critical points. The critical exponents of Heisenberg $O(4)$ model has been expected~\cite{PhysRevD.29.338}.  We choose the point ($T^\mathrm{C} \simeq193$ MeV, $\mu^\mathrm{C}=0$ MeV), which are on the second-order line and attracts much interests in lattice QCD, and calculate the mean field critical exponents at it. We display the critical exponents at this point in TABLE.~\ref{tabpc}. The critical exponent of order parameter $\beta$ agrees with the mean field theory prediction 1/2~\cite{PhysRevD.75.054026}, so $\gamma_T^\mathrm{C}$ is also expected to be 1/2, for the same reason mentioned for the critical exponents at the TCP.
We also calculate the critical exponents of some points on the second-order line. The result shown in Fig.~\ref{beta} demonstrates that most of the points on the second-order line share the same critical exponent. However as the point gets close to the TCP along the  second-order line, the critical exponent of the order parameter decreases continuously from its mean field value to the one at TCP. In other words, there is a crossover of different universality classes, although the TCP and the $O(4)$ line are both of second-order and their critical exponents are calculated in mean field approximation.

\begin{table}
\begin{center}
\caption{\label{tabtcp}Critical exponents at the TCP.}
\begin{tabular}{c c c c r @{.} l}\hline\hline
Quantity &  \hspace{3mm}Path    &  \hspace{3mm} Numerical result & \hspace{3mm}MF exponent \\\hline
 \multirow{2}{*}{$\beta^\mathrm{TCP}$}
         &$\rightarrow$   &  0.25$\pm$0.01                        & \multirow{2}{*}{1/4} \\
         &      $\uparrow$      &    0.25$\pm$0.01                              \\\hline
 \multirow{2}{*}{$\gamma_s^\mathrm{TCP}$}
 &$\rightarrow$   &  0.99$\pm$0.01                       & \multirow{2}{*}{1} \\
         &      $\uparrow$      &    1.00$\pm$0.01                              \\\hline
 \multirow{2}{*}{$\gamma_q^\mathrm{TCP}$}
         &$\rightarrow$   &  0.51$\pm$0.01                      & \multirow{2}{*}{1/2} \\
         &      $\uparrow$      &    0.51$\pm$0.01                         \\\hline

  \multirow{2}{*}{$\gamma_T^\mathrm{TCP}$}
          &$\rightarrow$   &  0.73$\pm$0.01                         & \multirow{2}{*}{---} \\
         &      $\uparrow$      &    0.74$\pm$0.01                               \\\hline
  \multirow{2}{*}{$\gamma_m^\mathrm{TCP}$}
          & $\rightarrow$   &  0.75$\pm$0.01                         & \multirow{2}{*}{---}  \\
         &      $\uparrow$      &    0.76$\pm$0.01                             \\\hline
  \multirow{2}{*}{$\gamma_n^\mathrm{TCP}$}
         &$\rightarrow$   &  0.48$\pm$0.01                         & \multirow{2}{*}{---} \\
         &      $\uparrow$      &    0.49$\pm$0.01                              \\\hline\hline

\end{tabular}
\end{center}
\end{table}

\begin{table}
\begin{center}
\caption{\label{tabpc} Critical exponents at the  critical point with zero $\mu$}.

\begin{tabular}{c c c  c r @{.} l}\hline\hline
Quantity    &  \hspace{3mm}Path   & \hspace{3mm} Numerical result & \hspace{3mm}MF exponent \\\hline
$\beta^\mathrm{C}$         &        $\uparrow$ &         0.50$\pm$0.01                         &   1/2  \\
$\gamma_s^\mathrm{C}$         &       $\uparrow$ &       1.01$\pm$0.01                         &  1   \\
$\gamma_T^\mathrm{C}$         &       $\uparrow$ &          0.51$\pm$0.01                         &   ---   \\\hline\hline
\end{tabular}
\end{center}
\end{table}

\section{SUMMARY}\label{sac}
In this paper, we have used two-flavor Nambu-Jona-Lasinio model to study the chiral phase transition at finite temperature and chemical potential in the cases of finite current quark mass as well as the chiral limit. We have compared the multi-solution region of Nambu and Wigner ones in the massive and massless quark cases in the phase diagram. The analysis of the multi-solution region is necessary, which provides us the exact first-order phase transition line, and then we can calculate the susceptibilities and corresponding critical exponents. The physical quantities, i.e., the effective quark mass and quark number susceptibility in the $T-\mu$ plane are presented for the massive and massless current quark case, respectively. We have used the susceptibilities to locate the CEP/TCP and studied the critical behaviors in the vicinity of them. This study reveals that all susceptibilities show the same critical behavior at CEP, and they are all found to be governed by a common critical exponent $\gamma$=2/3, for four different directions to approach the CEP. While in the chiral limit, the various susceptibilities show different features along the second-order transition line and so are the corresponding critical exponents. In addition, the critical exponent for order parameter $\beta^\mathrm{TCP}$ equals 1/4 at TCP, while $\beta^\mathrm{C}$ is 1/2 on the second-order line.
But the critical behavior in the vicinity of the points on the second phase transition line, belongs to the $O(4)$ universality class, with the mean field critical exponent for the order parameter $\beta^\mathrm{C}$=1/2. The different critical exponents of the TCP and the $O(4)$ line, indicates a change of different universality classes.
Last but not least, the mean field critical exponents is presented in this study, while the discussions beyond the mean field are also interesting, which no doubt deserves further study.

\acknowledgments
This work is supported in part by the National Natural Science Foundation of China (under Grants No. 11275097, No. 11475085, and No. 11535005), and the Jiangsu Planned Projects for Postdoctoral Research Funds (under Grant No. 1402006C).

\bibliographystyle{apsrev4-1}
\bibliography{luya}
\end{document}